 \documentclass[final,5p,times,twocolumn]{elsarticle}

\usepackage{epsfig}

\usepackage{amssymb}
\usepackage{amsmath}
\usepackage{array}    
\usepackage{amsthm}   
\usepackage{amsmath}  
\usepackage{thmtools} 
\usepackage{fancyvrb} 
\usepackage{amsfonts} 
\usepackage{tikz}
\usepackage{hyperref}
\usepackage{hhline}
\usepackage{rotating}
\usetikzlibrary{decorations.pathreplacing, positioning}
\usepackage{setspace}
\usepackage{caption}
\usepackage{algorithm}
\usepackage{algorithmic}
\usepackage{booktabs}
\usepackage[]{mathtools}
\usepackage{amssymb}
\usepackage{enumitem}
\usepackage{imakeidx}
\usepackage{mathrsfs}
\usepackage{tikz}
\usepackage{amsmath}
\usepackage{makecell}
\usepackage{tikz-3dplot}
\usepackage{graphicx}
\usepackage{epstopdf}
\usetikzlibrary{shapes.geometric, positioning, arrows.meta, fit, backgrounds, calc, quotes, matrix, shapes.arrows}
\usepackage{multirow}
\usepackage[numbers]{natbib}
\usepackage{float}

\newcommand{\N}{\mathbb{N}}

\newcommand{\R}{\mathbb{R}}

\newcommand{\s}{\mathbf{s}}

\newcommand{\abs}[1]{ \left| #1\right|}

\journal{Applied Acoustics}

\begin{document}

\begin{frontmatter}



\title{Reinforcement learning to design near-perfect acoustic absorbers}


\author[1]{M. Novelli}



\ead{mnovell@posgrado.upv.es}



\affiliation[1]{organization={Instituto Universitario de Matemática Pura y Aplicada (IUMPA), Universitat Politècnica de València (UPV)},
    addressline={Camí de Vera s/n}, 
    city={València},
    postcode={46022}, 
    country={Spain}}

\author[1]{R. Arnau}
\ead{roger.arnau@upv.es}

\author[1]{L.M. Garcia-Raffi}
\ead{lmgarcia@mat.upv.es}



\author[1]{V. Romero-García\corref{cor1}}
\ead{virogar1@mat.upv.es}
\cortext[cor1]{Corresponding author}



\begin{abstract}
Reinforcement learning (RL) is proposed as a strategy for the multi-objective design of Helmholtz resonators with near-perfect acoustic absorption. The approach formulates resonator design as a sequential decision-making problem in which geometrical parameters are iteratively updated to balance competing objectives, including resonator volume, resonance frequency, and acoustic reflection. A Q-learning algorithm based on a neural network is developed to explore the design space and identify non-dominated solutions, providing a Pareto front of optimal configurations. The acoustic response is evaluated using a transfer matrix method that incorporates viscothermal losses and acoustic end corrections. Three representative designs are subsequently investigated using full-wave finite element simulations, showing good agreement in their absorption and reflection spectra and validating the accuracy of the reduced-order model. Finally, one optimized resonator is fabricated by three-dimensional printing and experimentally characterized in an impedance tube. The measured absorption coefficient reaches approximately 95\% near the resonance frequency, in agreement with the numerical predictions. These results demonstrate the potential of reinforcement learning for the automated design of compact acoustic absorbers and establish a computationally efficient framework combining multi-objective optimization, numerical modelling, and experimental validation.
\end{abstract}


\begin{highlights}
\item Design of near-perfect acoustic absorbers with reinforcement learning techniques.
\item Q-learning technique based on neural networks for solving a multiobjective optimization problem.
\item Experimental validation near-perfect absorption in good agreement with the theoretical predictions.
\end{highlights}


\end{frontmatter}



\section{Introduction}
Perfect acoustic absorption can be achieved in linear lossy systems through the mechanism of critical coupling \cite{Yariv2000Universal, Chong2010Coherent, RomeroGarcia2016Perfect, RomeroGarcia2016ComplexFrequency, RomeroGarcia2020Design, Long2017Perfect, Achilleos2017NonHermitian, Qu2022AbsorptionMetamaterials, Merkel2015Control}, where intrinsic losses are exactly balanced by radiation leakage. Under this condition, incoming acoustic energy is neither reflected nor transmitted and is instead fully dissipated inside the absorber. In one port systems, perfect absorption occurs when the reflection coefficient vanishes, and, as transmission is suppressed by design, the absorption reaches unity \cite{Jimenez2016Ultrathin, LiAssouar2016, ZhangHu2016SinglePort, DampedResonance2019, Jimenez2018PerfectAbsorption, Cavalieri2020Graded, Boulvert2020Folded}. In two port systems, perfect absorption requires both reflection and transmission to vanish, which can be accomplished through multi-resonant designs \cite{Merkel2015Control,JimenezSciReps2017, Boulvert2022Asymmetric, Boulvert2022Compact, Meng2023Subwavelength}, allowing for one-directional perfect absorption, or by degenerate resonators \cite{Wei2014CoherentAbsorption, RomeroGarcia2020MirrorSymmetric, Yang2015DegenerateResonators} in symmetric configurations allowing for perfect absorption. In both cases, one and two ports, the perfect absorption arises when the internal (material or viscous) loss rate equals the external coupling (radiation) loss rate. At this balance point—known as critical coupling—the system impedance is perfectly matched to the surrounding medium. 

The design of resonators date from some decades \cite{Ingard1953AcousticResonators} but systems with perfect absorption based on critical coupling \cite{JimenezSciReps2017, Boulvert2020Folded, Boulvert2022Compact, Meng2023Subwavelength} are recent and more importantly are not obtained accidentally. They are the result of systematic optimization aimed at balancing radiation leakage and intrinsic dissipation while enforcing destructive interference at a given frequency. Sometimes this balance is typically achieved by adjusting resonator dimensions, neck lengths, cavity volumes, membrane tension, or the thickness and porosity of lossy layers. For broadband or multi-frequency absorption, optimization extends beyond single-resonator tuning. Usually multiple coupled resonances \cite{Boulvert2022Compact, Meng2023Subwavelength}, graded resonator arrays \cite{JimenezSciReps2017}, or hybrid structures combining different loss mechanisms are employed \cite{Boulvert2020Folded}. By distributing resonance frequencies and coupling strengths, absorption peaks can be merged or flattened. Numerical optimization techniques such as gradient-based methods \cite{Jimenez2016Ultrathin}, genetic algorithms \cite{JimenezSciReps2017}, bayesian techniques \cite{Li2026PhysicallyAdmissible}, or topology optimization \cite{Andersen2019ShapeOptimization, Tissot2020OptimalCavity}, are widely used to explore high-dimensional parameter spaces and identify configurations that achieve near-unity absorption over wider bands.

Recently, reinforcement learning (RL) has emerged as a promising optimization technique \cite{Shah2021Reinforcement, Gebrekidan2024Autonomous, Oh2026UnderwaterCoatings, Lu2026HelmholtzPorous, Wang2027ReinforcementLearning}, enabling the automated exploration of complex design spaces to identify geometrical configurations that achieve target acoustic performance without relying on conventional gradient-based optimization methods. In this work we design a RL algorithm which learns to solve a multiobjective problem to design acoustic materials made of Helmholtz resonators minimizing both the volume and the resonance frequency with impedance match conditions, or critical coupling conditions, in one port scattering systems. RL offers several distinct advantages for the design of perfect acoustic absorbers because it does not rely on labeled datasets. RL can explore the design space autonomously by interacting directly with numerical models or experiments, reducing the burden of data generation. Its formulation as a sequential decision-making process allows complex designs to be constructed through incremental parameter updates, which is particularly well suited to resonant absorbers where small changes can strongly affect coupling and loss balance. RL can naturally handle mixed discrete and continuous design variables and readily incorporate multiple, potentially competing objectives through carefully defined reward functions. Moreover, reinforcement learning does not require gradient information, making it applicable to non-smooth, non-convex, or black-box optimization problems commonly encountered in acoustic and elastic metamaterial design. 

\section{Acoustic Modeling}
Let us consider an acoustic problem under the hypothesis of plane-wave propagation. Generally, a problem of this type can be reduced to a one-dimensional one, even in complex cases. There are many techniques to solve this kind of problems analytically, but we will focus on the transfer matrix method (TMM) \cite{Jimenez2021TMM} that is fitted to the geometry we want to solve here. In this work we will use the time harmonic convention of the type $e^{i\omega t}$. We will apply the TMM to find the resonance frequency in the context of Helmholtz resonators as well as the scattering coefficients of the system. In addition to assuming plane-wave propagation, the model assumes that the materials are isotropic and homogeneous. In general, these hypothesis are preserved in the range of frequencies of our analysis as the geometries are selected accordingly to this purpose. Moreover, we will consider that the amplitude of the waves is small enough to be in the linear regime and to consider that the system is reciprocal.

In this work we consider a Helmholtz Resonator made of a cylindrical neck and a cavity. On one side it is connected to the main-waveguide through the neck, while on the other side it is closed. The walls of the system are considered acoustically rigid, i.e., with a Neumann boundary condition. Viscothermal losses are considered inside each part of the Helmholtz resonator by using the effective properties derived by Stinson \cite{Stinson1991}, as described in the next subsection.



\subsection{Viscothermal losses and length corrections}
In the lossy case, we refer to the case in which viscothermal losses are considered and the acoustic energy can be transferred to thermal one, so dissipation is possible. In this case the impedance and the bulk modulus are frequency dependent and complex. To compute the effective parameters is useful to define the following complex and frequency dependent functions
\begin{eqnarray}
    G_{\rho}\left(\omega\right) & = &\sqrt{-\dfrac{i\omega \rho_0}{\eta}},\\
    G_{K}\left(\omega\right) & = &\sqrt{-\dfrac{i\omega \text{Pr} \rho_0 }{\eta}}.
\end{eqnarray}
Here, $\omega$ is the angular frequency, $\rho_0$ is the air density, $\eta$ is the dynamic viscosity, and $\mathrm{Pr}$ is the Prandtl number.

Let us now consider a cavity with a circular section of radius $r$ (see Fig.~\ref{fig:fig2}). The density and the bulk modulus take the following form
\begin{eqnarray}
    \rho\left(\omega\right) & =& \rho_0 \left(1 - \dfrac{2}{rG_{\rho}\left(\omega\right)}\dfrac{J_1\left(r G_{\rho}\left(\omega\right)\right)}{J_0\left(r G_{\rho}\left(\omega\right)\right)}\right)^{-1},\\
    K\left(\omega\right) & = &K_0 \left(1 + \dfrac{2\left(\gamma-1\right)}{rG_{K}\left(\omega\right)}\dfrac{J_1\left(r G_{K}\left(\omega\right)\right)}{J_0\left(r G_{K}\left(\omega\right)\right)}\right)^{-1},
\label{cylindrical}
\end{eqnarray}
where $J_0$ and $J_1$ are the Bessel functions of the first kind of order 0 and 1, respectively. Here, $K_0$ is the bulk modulus of air and $\gamma$ is the ratio of specific heats. These frequency-dependent quantities allow us to calculate, when considering thermoviscous losses, the normalized acoustic impedance and the wavenumber:
\begin{equation}
    Z'\left(\omega\right) = \frac{\sqrt{\rho\left(\omega\right)K\left(\omega\right)}}{S}, \qquad k\left(\omega\right) = \omega \sqrt{\dfrac{\rho\left(\omega\right)}{K\left(\omega\right)}},
\end{equation}
where $S=\pi r^2$ is the cross-sectional area of the considered cylindrical duct.


\subsection{Acoustic impedance}
\begin{figure*}
    \centering
    \includegraphics[]{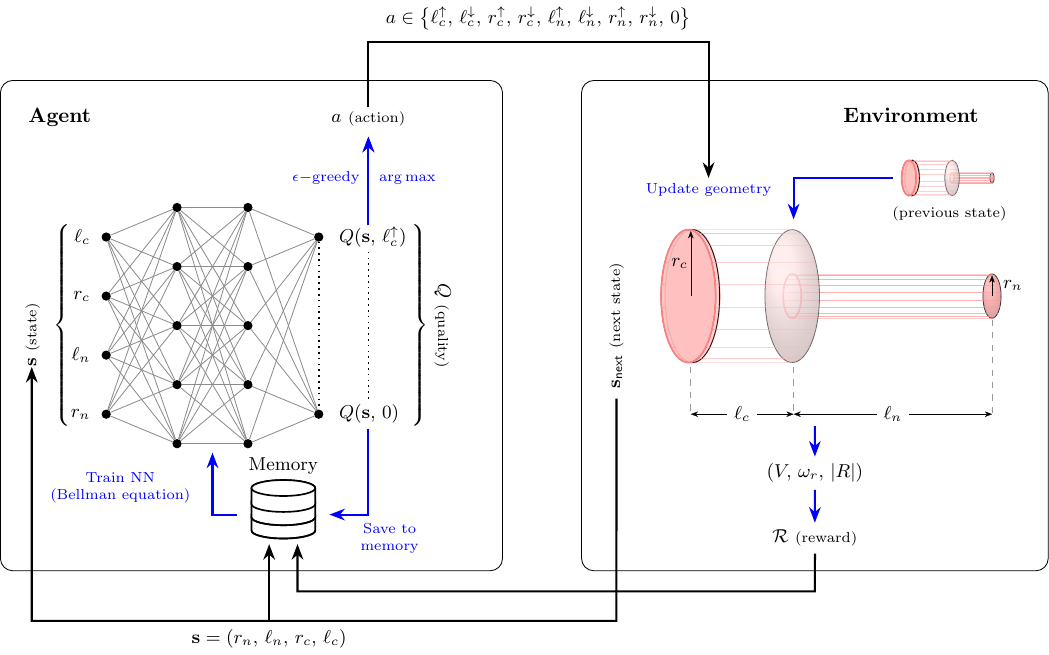}
    \caption{Sketch for the reinforcement learning and Q-learning.}
    \label{fig:fig2}
\end{figure*}

Let denote by $r_n$ and $\ell_n$ ($r_c$ and $\ell_c$) the radius and the length of the neck (cavity) respectively (Fig.~\ref{fig:fig2}). We may relate the sound pressure, $p$, and the flux, $\mathcal{V}_z=Sv_z$, where $S$ is the cross section and $v_z$ is the particle velocity, at the beginning of the duct, $z=0$, and at the end, $z=\ell_c+\ell_n \eqqcolon L$, through a total transfer matrix. We consider a length correction (for further details, see \cite{vicent}, p. 129), $\Delta \ell_2$, at the beginning of the neck (at $z=0$) and another length correction, $\Delta \ell_1$, at the end (at $z=\ell_n$). 

\begin{align}
&\begin{pmatrix}
    p\\ \mathcal{V}_z
\end{pmatrix}_{z=0}  =\nonumber\\&
\begin{pmatrix}
    \cos{\left(k_n \Delta \ell_2\right)} & iZ'_n\sin{\left(k_n \Delta \ell_2\right)}\\
    i\dfrac{1}{Z'_n}\sin{\left(k_n \Delta \ell_2\right)} & \cos{\left(k_n \Delta \ell_2\right)}
\end{pmatrix}\nonumber\\&
\begin{pmatrix}
 \cos{\left(k_n \ell_n\right)} & iZ'_n\sin{\left(k_n \ell_n\right)}\\
    i\dfrac{1}{Z'_n}\sin{\left(k_n \ell_n\right)} & \cos{\left(k_n \ell_n\right)} 
\end{pmatrix}
 \begin{pmatrix}
    \cos{\left(k_n \Delta \ell_1\right)} & iZ'_n\sin{\left(k_n \Delta \ell_1\right)}\\
    i\dfrac{1}{Z'_n}\sin{\left(k_n \Delta \ell_1\right)} & \cos{\left(k_n \Delta \ell_1\right)}
\end{pmatrix}\nonumber\\&
\begin{pmatrix}
    \cos{\left(k_c \ell_c\right)} & iZ'_c\sin{\left(k_c \ell_c\right)}\\
    i\dfrac{1}{Z'_c}\sin{\left(k_c \ell_c\right)} & \cos{\left(k_c \ell_c\right)}
\end{pmatrix}
\begin{pmatrix}
    p\\ \mathcal{V}_z
\end{pmatrix}_{z=L}=\mathcal{T}\begin{pmatrix}
    p\\ \mathcal{V}_z
\end{pmatrix}_{z=L}.
\label{TTM_corr}
\end{align}
From Eq.~(\ref{TTM_corr}) the normalized acoustic input impedance at the entrance of the resonator can be obtained as 
\begin{equation}
    Z'_r =\frac{p(z=0)}{\mathcal{V}_z(z=0)}= \frac{\mathcal{T}_{1,1}}{{\mathcal{T}_{2,1}}},
\end{equation}
since $\mathcal{V}_z\left(z=L\right)=0$ due to the rigid wall at the end of the resonator. If one assumes that $k_n \Delta \ell_1 \ll 1$ and $k_n \Delta \ell_2 \ll 1$, a first order approximation of this normalized acoustic input impedance, $Z_r'\coloneqq\dfrac{p}{\mathcal{V}_z}$ at $z = 0$, is
\begin{equation}
\resizebox{0.48\textwidth}{!}{$
\displaystyle
Z'_r = -i Z'_n \dfrac{1 - k_n \Delta l \left( \dfrac{Z'_n}{Z'_c}\tan{\left(k_c \ell_c\right)} + \tan{\left(k_n \ell_n\right)} \right) - \dfrac{Z'_n}{Z'_c} \tan{\left(k_n \ell_n\right)} \tan{\left(k_c \ell_c\right)} }{\tan{\left(k_n \ell_n\right)} + k_n \Delta l \left( 1 - \dfrac{Z'_n}{Z'_c} \tan{\left(k_n \ell_n\right)} \tan{\left(k_c \ell_c\right)} \right) + \dfrac{Z'_n}{Z'_c} \tan{\left(k_c \ell_c\right)} } $}.
\label{correction}
\end{equation}
 
\subsubsection{Numerical end-corrections in the cylindrical case}
For a cylindrical geometry, with the resonator connected in series with a cylindrical main-waveguide of radius $r_{mw}$, let
\begin{equation}
    \varepsilon_1 \coloneqq \frac{r_n}{r_c}, \quad
    \varepsilon_2 \coloneqq \frac{r_n}{r_{mw}}.
\end{equation}
The following numerical approximations for the length corrections are considered:
\begin{equation}
    \Delta \ell_i \approx
    0.82 \left(1 - 1.35 \varepsilon_i + 0.31 \varepsilon_i^3\right) r_n,
    \quad i = 1,2.
\end{equation}
For $i=1$, $\Delta \ell_1$ refers to the length correction to be considered at the neck--cavity interface, whereas for $i=2$, $\Delta \ell_2$ refers to the correction at the neck--main-waveguide interface. These approximations are accurate for 
\begin{equation}
    \varepsilon_i \leq 0.55, \quad i =1,2.
    \label{rr_055}
\end{equation}
Further details on these approximations can be found in \cite{Dubos, Dalmont1998Measurement, Dalmont2001Radiation}. The resonator shape optimization process was carried out considering only resonators whose geometry satisfies Eq.~\eqref{rr_055}.


\subsection{Resonance frequency and numerical search}
The search for the resonance frequency of the Helmholtz resonator becomes equivalent to find a root of the imaginary part of the complex acoustic input impedance of the resonator. Such a root cannot be found, in general, analytically; therefore, it is necessary to use numerical methods. In particular, the secant method was used. 

Consider an interval $I\subseteq \R$ and a function $f:I \to \R,\; x \mapsto f\left(x\right)$, whose root is to be found. The secant method allow a sequence of points to be generated which, under certain conditions, converges to the desired root starting from one or more suitably chosen initial points. The performance of the method is strongly dependent on these initial choices. In particular, it requires two initial points, $x_0$ and $x_1$, and generates the new iterates through the formula
\begin{equation}
    x_{i} = x_{i-1} - \dfrac{x_{i-1} - x_{i-2}}{f\left(x_{i-1}\right) - f\left(x_{i-2}\right)}f\left(x_{i-1}\right), \quad i = 2, 3, \ldots 
\end{equation}
The secant method does not require knowledge of the derivative of the function (which is required, for example, in Newton's method). 
Moreover, this method has a superlinear local convergence rate.

\subsubsection{Numerical computation of the resonance frequency}
In our case, when losses are neglected, the normalized acoustic input impedance $Z'_r\left(\omega\right)$ takes only imaginary values ($\rho\left(\omega\right) \equiv \rho_0$ and $K\left(\omega\right) \equiv K_0$). Therefore, by neglecting the imaginary unit, it can be regarded as a real-valued function.  When losses are introduced, however, the function $Z'_r\left(\omega\right)$ becomes complex-valued and frequency-dependent. Since its imaginary part cannot be expressed in analytical form, its derivative cannot be calculated either: that's why, in order to find its first positive root, a derivative-free root-finding method had been used.

For the choice of the initial points it is assumed that the function $\Im\left(Z'_r\left(\omega\right)\right)$ possesses the following properties, regardless of whether end-corrections are taken into account:
\begin{itemize}
    \item There is an asymptote in $\omega = 0$: recalling that $k\left(\omega\right) = \omega\dfrac{\rho\left(\omega\right)}{K\left(\omega\right)}$, this statement can be easily verified considering that $k_n\left(0\right) = k_c\left(0\right) = 0$. This implies that the denominator vanishes when $\omega = 0$ in all the cases, considering or not the end-corrections. Moreover, the function is negative in a right neighbourhood of 0.
    \item There exists a point $\omega_r > 0$ such that $\Im\left(Z'_r\left(\omega_r\right)\right) = 0$. The smallest positive value satisfying this condition corresponds to the resonance frequency.
    \item There exists $\widetilde{\omega} > \omega_r$ such that $\Im\left(Z'_r\left(\omega\right)\right)$ is concave for $\omega \in \left(0,\,\widetilde{\omega}\right)$. 
    \item $\Im\left(Z'_r \left(\omega\right)\right)$ is an increasing function for $\omega \in \left(0,\,\widetilde{\omega}\right)$.
    \item Fixing the parameters of the Helmholtz resonator, the root encountered in the case with end-correction is smaller than that in the no end-corrections case.
\end{itemize}
These assumptions are valid for an ideal damped and driven harmonic oscillator and, given the analogy between this model and the one describing a Helmholtz resonator, they can also be assumed to hold in the case under consideration. Under these assumptions, convergence of the secant method to the first positive root is guaranteed by choosing the initial points in the interval where the function is negative and concave, i.e., for $\omega \in \left(0,\, \omega_r\right)$. In particular, the following initial points were chosen:
\begin{equation}
    \omega_0 = 1,\; \omega_1 = 2.
\end{equation}
If one of these values was found to be positive, it was progressively halved until the negativity condition was satisfied.

\subsection{Reflection coefficient}

We consider the resonator to be connected at one end of another cylindrical tube (the main waveguide) with radius $r_{mw}$. The cross-sectional area of the main waveguide is $S_{mw} = \pi r_{mw}^2$. The radius $r_{mw}$ is assumed to be sufficiently large that viscothermal losses in this tube can be neglected (hence, $Z_{mw}'$ is constant). The density and sound speed inside the main waveguide are those of air (or of the medium under consideration), namely $\rho_{mw} = \rho_0$ and $c_{mw} = c_0$.The pressure reflection coefficient at the main waveguide–resonator interface is then
\begin{equation}
R\left(\omega\right) = \dfrac{Z'_r\left(\omega\right) - Z_{mw}'}{Z'_r\left(\omega\right) + Z_{mw}'} .
\end{equation}
The absorption coefficient is then given by $\alpha\left(\omega\right) = 1 - \abs{R\left(\omega\right)}^2 $. Perfect absorption will be achieved when the impedance matching condition at the resonance frequency, $Z'_r\left(\omega_r\right) = Z_{mw}'$, is satisfied. From now on, we will use the shorthand $R = R \left( \omega_r \right)$, since we are only interested in its value at the resonance frequency.

\section{Reinforcement Learning}
Physics often involves optimization problems that are typically non-convex and difficult to solve using traditional analytical or numerical methods. For this reason, it is necessary to introduce more advanced tools that can achieve results by bypassing the limitations of traditional models. This is where artificial intelligence algorithms, and in particular machine learning, come into play (see \cite{Gebr2024}, \cite{Mitchell1997} and \cite{Yu20}). These algorithms are capable of learning to optimize the aforementioned problems simply by leveraging experience from data.

A particularly effective tool in these situations is RL, a class of machine learning algorithms that ``learn'' to find optimal solutions to problems through a reward system that evaluates the ``quality'' of the decisions made. The idea is to mimic human learning, which, through interaction with the surrounding environment, learns to choose the best action to take in various contexts (for further information see \cite{SuttonBarto2018} and \cite{Zhao2024}).

The two main entities in RL are the \textit{environment} and an \textit{agent} that makes decisions influencing the environment. Each interaction between the agent and the environment is called a \textit{step}, while the situation presented by the environment at each step is called a \textit{state}. The agent, observing the current state of the environment, takes a decision called an \textit{action}. The criterion for choosing actions, called the agent’s \textit{policy} and denoted by $\pi$, is refined through experience and guided by the reception of a \textit{reward} that either rewards or penalizes the agent’s work at each step.

In this way, starting from an initial state, the agent performs a sequence of steps until reaching a final state. This sequence is called an \textit{episode}. Through numerous episodes, each consisting of several steps, the algorithm hopefully learns to navigate optimally in any situation.

RL allows solving, in particular, multi-objective problems like the one we are dealing with. To see other examples of RL applied to multi-objective problems see \cite{Hayes2022} and
\cite{VanMoffaert2014}.

\subsection{Model Construction}
Let $\mathcal{S}$ be the set of possible states and $\mathcal{A}$ the set of actions that the agent can take. The policy $\pi$ is defined as follows:
\begin{align}
    \pi : \mathcal{S} & \to \mathcal{A}\nonumber\\
    \s & \mapsto \pi\left(\s\right).
\end{align}
where $\pi\left(\s\right)$ should be the best action that can be taken from state $\s$. The function $\pi$ is not known, and the objective is to learn it.
If we assume that the state transitions are deterministic, as in our case, the environment can be modeled by the following transition function.
Given a state $\s \in \mathcal{S}$ and a set of possible actions $a \in \mathcal{A}$, it is possible to define a function that maps each state-action pair to the next state $\s_{\text{next}}$:
\begin{align}
    \phi : \mathcal{S} \times \mathcal{A} & \to \mathcal{S}\nonumber\\
    \left(\s,\, a\right) & \mapsto \phi\left(\s,\, a\right) = \s_{\text{next}}.
\end{align}
However, the agent does not have prior knowledge of $\phi$.
The strategy by which, in our case, the agent selects the optimal action is called \textit{Q-learning} and was introduced in 1989 by Christopher Watkins in \cite{Watkins1989}. However, in this work the version proposed in \cite{Mnih2013} and used in \cite{Arnau2023} will be employed.

At each step, it is not only important to maximize the reward received at that step, but also those that will be received in the future. Let $R_n\left(\s\right)$, $n\in\mathbb{N}$, be the reward received after $n$ steps taken starting from the current state $\s$ using policy $\pi$. The value of a state $\s$, according to the policy $\pi$, is defined as the function
\begin{align}
    V_{\pi}: \Omega & \to \mathbb{R}\nonumber\\
    \s & \mapsto \sum_{n=0}^{+\infty} \gamma^n \mathcal{R}_n\left(\s\right),
\label{V_pi}
\end{align}
where $0<\gamma<1$ is the \textit{discount factor}, which serves to prioritize the impact of rewards received sooner over those received later when computing the value of the state. Note that if the reward values $\mathcal{R}_n$, $n \in \mathbb{N}$, are bounded, then the series is convergent (its modulus is dominated by a multiple of a convergent geometric series). The goal is to find a policy $\pi$ that maximizes the value $V_{\pi}\left(\s\right)$ for every $\s \in \mathcal{S}$. By manipulating formula \eqref{V_pi}, we obtain
\begin{align}
    V_{\pi}\left(\s\right) & = \mathcal{R}_0\left(\s\right) + \gamma \cdot \sum_{n=0}^{+\infty} \gamma^n \mathcal{R}_{n+1}\left(\s\right)\nonumber\\
    & = \mathcal{R}_0\left(\s\right) + \gamma\cdot \sum_{n=0}^{+\infty} \gamma^n \mathcal{R}_n\left(\s_{\text{next}}\right)\nonumber\\
    & = \mathcal{R}_0\left(\s\right) + \gamma \cdot V_{\pi}\left(\s_{\text{next}}\right).
\end{align}
The search for the optimal policy is carried out by defining a “quality” function
\begin{align}
    Q: \Omega \times \mathcal{A} & \to \mathbb{R}\nonumber\\
    \left(\s,\,a\right) & \mapsto Q\left(\s,\,a\right) = \nonumber\\& \sup\left\{V_{\pi}\left(\s\right)\ \middle|\ \pi: \Omega\to \mathcal{A} \text{ is a policy and } \pi\left(\s\right) = a\right\}.
\end{align}
The reward $\mathcal{R}_0\left(\s\right)$ is clearly the reward obtained after taking a certain action $a \in \mathcal{A}$, so we can denote it by $\mathcal{R}\left(\s,\,a\right) \coloneqq \mathcal{R}_0\left(\s\right)$. We obtain:
\begin{align}
    &Q\left(\s,\,a\right) =\nonumber\\& 
    \mathcal{R}\left(\s,\,a\right) + \gamma \cdot \sup\left\{V_{\pi}\left(\s_{\text{next}}\right)\ \middle|\ \pi: \Omega\to \mathcal{A} \text{ is a policy and } \pi\left(\s\right) = a\right\}\nonumber\\& 
     \approx \mathcal{R}\left(\s,\,a\right) + \gamma \cdot \sup_{a'\in \mathcal{A}} Q\left(\s_{\text{next}},\, a'\right).
\label{Q_approx}
\end{align}
From this approximation, the final policy will be
\begin{equation}
    \pi\left(\s\right) = \arg\max \left\{Q\left(\s,\,a\right) \ \middle|\ a \in \mathcal{A}\right\}.
\end{equation}
With this definition, we also obtain
\begin{equation}
    V_{\pi}\left(\s\right) = \max \left\{Q\left(\s,\,a\right) \ \middle|\ a \in \mathcal{A}\right\} = Q\left(\s,\,\pi\left(\s\right)\right).
\end{equation}
The initial objective, namely finding the optimal policy $\pi$, thus translates into the search for the function $Q$. This can be done using a neural network. The model described so far would lead to a neural network with two inputs: the state $\s$ and the action $a$. However, it is more convenient to design a network that takes only the state $\s$ as input and returns the vector $\left[Q\left(\s,\,a\right)\right]_{a\in \mathcal{A}}$, i.e., the vector containing all the $Q$ values expected for each pair $\left(\s,\,a\right)$ with $\s$ fixed. The action chosen will be the one that maximizes the value of $Q\left(\s,\,\cdot\right)$. Formula \eqref{Q_approx} is called Bellman optimality equation for the state-action value function (see \cite{Bellman1957}). 

In order for the formula \eqref{Q_approx} to be satisfied, the quadratic error
\begin{equation}
    \left(Q\left(\s,\,a\right) - \left[ \mathcal{R}\left(\s,\,a\right) + \gamma \cdot \sup_{a'\in \mathcal{A}} Q\left(\s_{\text{next}},\, a'\right)\right]\right)^2
    \label{error}
\end{equation}
must be as small as possible. The weights of the network should ideally be updated using all the experiences stored in memory, considering the mean of errors like \eqref{error}, that is considering the \textit{Mean Squared Error (MSE)} loss function. However, to reduce computational cost and accelerate training, it is common practice to sample a batch of data from the available memory at each training step, rather than using the entire memory.

Relying solely on actions suggested by the $Q$ function computed by the network may prevent the exploration of certain areas of the action space.  
For this reason, the $\epsilon$-greedy strategy is implemented, according to which:
\begin{itemize}
    \item With probability $1 - \epsilon$, the action suggested by the $Q$ function learned by the neural network is selected;
    \item With probability $\epsilon$, an action $a \in \mathcal{A}$ is randomly selected.
\end{itemize}
In this way, a balance is achieved between \textit{exploitation}, i.e., leveraging accumulated experience, and \textit{exploration}, i.e., searching for potentially better alternatives. Moreover, since the network is not yet trained at the beginning, it is common to start with a relatively high value of $\epsilon$, denoted $\epsilon_0$ (for example, $\epsilon_0 = 1$), and gradually decrease it as the agent gains experience. In this way, the agent initially performs extensive exploration and, over time, increasingly follows the suggestions of the neural network. To prevent the agent from completely stopping exploration, a minimum value $\epsilon_{\min}$ is set, below which $\epsilon$ cannot decrease. Therefore, the update of the $\epsilon$ value at each step is performed as follows:
\begin{equation}
    \epsilon \leftarrow \max \left\{ \epsilon_{\min},\ \epsilon \cdot \epsilon_{\text{decay}} \right\},
    \label{aggiornamento_epsilon}
\end{equation}
where $0 < \epsilon_{\text{decay}} < 1$ is the decay parameter for $\epsilon$. In particular, in order to reach the value $\epsilon_{\min}$, $n$ steps are required, where $n \in \mathbb{N}$ is the smallest natural number such that
\begin{equation}
    \epsilon_{\min} \leq \epsilon_0 \cdot \left(\epsilon_{\text{decay}}\right)^n. 
\end{equation}
This occurs when
\begin{equation}
    n = \Bigg\lceil \log_{\epsilon_{\text{decay}}} \left(\dfrac{\epsilon_{\min}}{\epsilon_0}\right) \Bigg\rceil.
    \label{epsilon_dec}
\end{equation}

Given the number of episodes $n_{\text{ep}}$ and the number of iterations $n_{\text{it}}$, the reinforcement learning algorithm that arises from the previous discussion is Algorithm \ref{alg:reinf}.

\begin{algorithm}
\begin{algorithmic}
\FOR{$i=1$ to $n_{\text{ep}}$}
  \STATE $\s = $ randomly chosen initial state;
  \FOR{$j=1$ to $n_{\text{it}}$}
    \STATE Generate a random number $r_a \in \left[0,1\right]$;
    \IF{$r_a < \epsilon$}
      \STATE $a = $ randomly selected action;
    \ELSE
      \STATE $a = \arg\max_{a \in \mathcal{A}} Q\left(\s,\,a\right)$;
    \ENDIF
    \STATE $\s_{\text{next}} = \phi\left(\s,\, a\right)$;
    \STATE $\mathcal{R}\left(\s,\, a\right) = $ reward;
    \STATE Store $\left(\s,\, a,\, \mathcal{R}\left(\s,\, a\right),\, \s_{\text{next}}\right)$ in memory;
    \STATE Train the neural network;
    \STATE $\s = \s_{\text{next}}$;
  \ENDFOR
\ENDFOR
\end{algorithmic}
\caption{Algorithm related to the reinforcement learning process}
\label{alg:reinf}
\end{algorithm}

\subsection{Reinforcement learning Applied to the Optimization of a Helmholtz Resonator Shape}

In our case, the multi-objective problem we aim to solve is to design a Helmholtz Resonator with the smallest possible volume, $V$, the lowest possible resonance frequency, $\omega_r$, and the smallest absolute value of the reflection coefficient $|R|$. The volume of the resonator is given by the sum of the volumes of the neck and the cavity:
\begin{equation}
    V = V_n + V_c = S_n \ell_n + S_c \ell_c,
\end{equation}
where $\ell_n$ and $\ell_c$ are the lengths of the neck and the cavity, respectively, while $S_n = \pi r_n^2$ and $S_c = \pi r_c^2$ are their respective cross-sectional areas. The three goals are conflicting, since a decrease in volume is usually associated with an increase in the resonance frequency and increase in the reflection coefficient. Therefore, the objective is to obtain a Pareto front from the results provided by a RL algorithm based on a neural network that learns the $Q$-function, and to extract an optimal solution from it (using the minimum distance criterion from the ideal vector in the $\left(V,\omega, |R|\right)$ space).

Let $n_p$ denote the number of parameters that define the resonator: clearly $n_p = 4$ for slits or cylindrical ducts, while $n_p = 6$ for a rectangular section. Given a modification rate $\delta$ for the parameters (in our case $\delta = 10^{-4}\,\mathrm{m}$), the possible actions the agent can take at each step are:
\begin{itemize}
    \item Increase one of the $n_p$ parameters of the resonator by an amount $\delta$.
    \item Decrease one of the $n_p$ parameters of the resonator by an amount $\delta$.
    \item Do nothing (this option is mainly chosen when an optimal configuration of the parameters has been reached).
\end{itemize}
Thus, there are $2\cdot n_p + 1$ possible actions that the agent can choose from. This means that the network represents a function $N:\R^{n_p} \to \R^{2\cdot n_p + 1}$. As previously described, the choice of the best action is made with probability $1 - \epsilon$ according to the suggestions from the $Q$-function computed by the network, and with probability $\epsilon$ in a random manner. The value of $\epsilon$ varies according to the strategy described in Eq.~\eqref{aggiornamento_epsilon}; specifically, we set $\epsilon_0 = 1$, $\epsilon_{\text{decay}} = 0.99999$, and $\epsilon_{\min} = 0.2$. Using equation \eqref{epsilon_dec}, it follows that the value $\epsilon_{\min}$ is reached in $n$ steps, where
\begin{equation}
    n = \Bigg\lceil \log_{0.99999} \left(\dfrac{0.2}{1}\right) \Bigg\rceil = 160943.
    \label{eps_greedy}
\end{equation}

The parameter space has been restricted to ensure the parameters never become less than or equal to zero, nor excessively large. This guarantees that the results make both physical and practical sense. In particular, for the case of cylindrical ducts, we consider vectors $\left[\ell_c,\, r_c,\, \ell_n,\, r_n\right] \in \Omega$ where 
\begin{equation}
\begin{split}
    \Omega = & \left\{\left[\ell_c,\, r_c,\, \ell_n,\, r_n\right] \in \left[0.01,\, 0.08\right] \times \left[0.0036, 0.04\right]\times \right.\\
    & \times \left[0.01,\, 0.04\right]\times \left[0.002, \, 0.022\right] \text{ and }\left. r_n \leq 0.55 \cdot r_c\right\}.
\end{split}
\end{equation}
The constraint $r_n \leq 0.55 \cdot r_c$ is related to the validity of the approximation in Eq.~\eqref{rr_055}. Since the numerical simulations were performed considering a main waveguide with radius $r_{mw} = 0.05\,\mathrm{m}$, Eq.~\eqref{rr_055} is always satisfied at the transition between the main waveguide and the resonator's neck. The choice of this waveguide radius is dictated by the geometry of the experimental apparatus used to validate the numerical results. In order to obtain a broad understanding of the entire search space, the initial geometry of each episode is sampled from a uniform distribution within the domain $\Omega$. Preliminary experiments were carried out by always, or almost always, starting from a fixed point. However, this led to overfitting and to a lower diversity in the geometry of the optimal solutions. For this reason, it was decided to always start from different points, although this requires a larger number of episodes and/or iterations to allow the neural network to learn effectively. 

\begin{figure*}
    \centering
    \includegraphics[width=\linewidth]{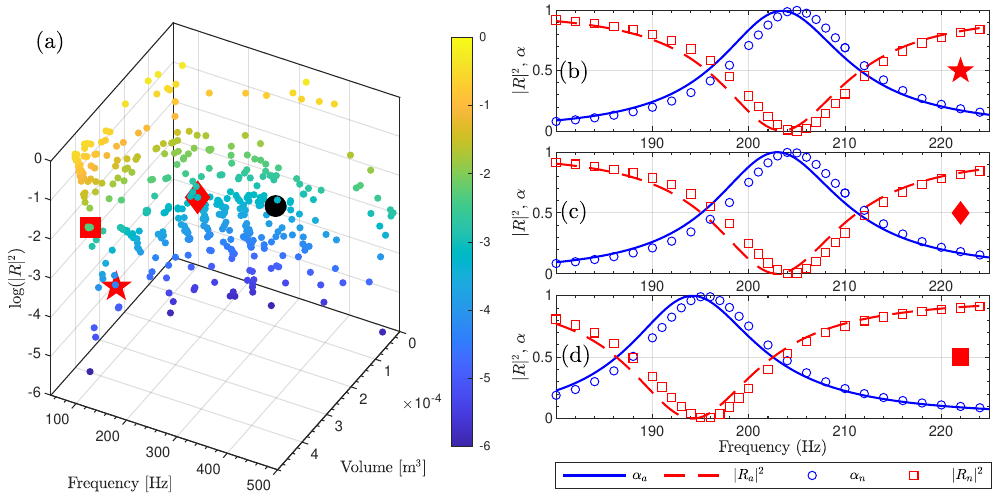} 
    \caption{(a) Pareto front obtained through RL. Red symbols correspond to the resonators chosen to test their performance numerically in (b), (c) and (d). Blakc symbol correspond to the resonator selected for 3D printing and experimental characterization. Their geometric parameters are shown in Tab.~\ref{tab:values}. (b-d) show the full wave numerical simulations (symbols) with the TMM predictions (lines). The resonator corresponding to the black symbol is analyzed in Fig.~\ref{fig:fig4}.}
    \label{fig:fronte_3D}
\end{figure*}

If, during the optimization process, the selected action results in a geometry outside the domain $\Omega$, the geometry is adjusted as follows:
\begin{itemize}
    \item If the action acts on $l_c$ or $l_n$, the corresponding parameter is projected onto the boundaries of the domain.
    \item If the action acts on $r_n$, two cases are considered. If the new radius is smaller than the minimum allowable neck radius, denoted by $r_{n,\min}$, then
    \begin{equation}
        r_n \leftarrow r_{n,\min}.
    \end{equation}
    If, at the current step, $r_n > 0.55 \cdot r_c$, then
    \begin{equation}
        r_n \leftarrow 0.55 \cdot r_c.
    \end{equation}
    \item If the action acts on $r_c$, two cases are considered. If the new radius is larger than the maximum allowable cavity radius, denoted by $r_{c,\max}$, then
    \begin{equation}
        r_c \leftarrow r_{c,\max}.
    \end{equation}
    If, at the current step, $r_n > 0.55 \cdot r_c$, then
    \begin{equation}
        r_c \leftarrow \frac{r_n}{0.55}.
    \end{equation}
\end{itemize}
Due to the precision limitations of the instrument used for the 3D printing of the resonators (see Subsection~\ref{experimental_results}), the values of the four geometric parameters involved are always rounded to the fourth decimal place, consistently with the value of $\delta$. 

Given the value $\delta = 10^{-4}\,\mathrm{m}$, it can be shown that at most $1564$~steps are required to move between any two points in~$\Omega$. To ensure that the algorithm can reach any state within the domain during an episode, the maximum number of iterations per episode is set to $n_{\text{it}} = 2000$. This choice provides a sufficient number of iterations to reach any point in the domain while also allowing for non-direct trajectories, which may arise when the algorithm explores the domain rather than moving directly towards its target. Additionally, the number of experiments is set to $n_{\text{exp}} = 750$. A high number of experiments is necessary to allow the algorithm to acquire sufficient knowledge on how to best act from any state. Furthermore, to encounter a significant number of Pareto-optimal solutions—given that the Pareto front is three-dimensional (involving three objective functions)—a vast number of optimal points is required to properly characterize the Pareto surface. At most $750 \times 2000 = 15\cdot 10^5$ steps will be taken in total (actually fewer, since, as described in Subsubsection \ref{sub_rewards}, some episodes are terminated early in case of prolonged stalemates). In any case, as seen in Eq.~\ref{eps_greedy}, the probability of randomly choosing the action to perform will be decreased over the first 160943 steps. 

\begin{table}
    \centering
    \begin{tabular}{|c|ccc|}
        \hline
         
        & $\mathcal{R}_V$
        & $\mathcal{R}_{\omega_r}$
        & $\mathcal{R}_R$\\
        \hline
        $\downarrow$ & 5 & 4 & 6\\ 
        $=$ & 0 & 0 & 0\\
        $\uparrow$ & -5 & -4 & -6\\
        \hline
    \end{tabular}
    \caption{Reward values associated with the variation of each objective function.}
    \label{tab:rewards}
\end{table}

\begin{table*}
\centering
\begin{tabular}{c|ccccc|c|ccc|}
\cline{2-10}
& \multicolumn{5}{c|}{\textbf{Geometric parameters}} & \textbf{Method} & \multicolumn{3}{c|}{\textbf{Acoustic parameters}} \\
\cline{2-10}
& $l_c\,[\mathrm{m}]$ & $r_c\,[\mathrm{m}]$ & $l_n\,[\mathrm{m}]$ & $r_n\,[\mathrm{m}]$ & $V\,[\times 10^{-5}\,\mathrm{m}^3]$ &  & $\omega_r\,[\mathrm{rad/s}]$ & $f_r\,[\mathrm{Hz}]$ & $\abs{R\left(\omega_r\right)}^2$\\
\cline{2-10}
\multirow{2}{*}{\textcolor{red}{{\Large$\star$}}}
& \multirow{2}{*}{0.0177} & \multirow{2}{*}{0.0367} & \multirow{2}{*}{0.0399} & \multirow{2}{*}{0.0040} & \multirow{2}{*}{$7.6901$} & TMM & 1278 & 203 & $3.97 \cdot 10^{-3}$\\
& & & & & & FEM & 1288 & 205 & $6.45\cdot 10^{-4}$ \\
\cline{2-10}
\multirow{2}{*}{\textcolor{red}{$\blacklozenge$}}
& \multirow{2}{*}{0.0234} & \multirow{2}{*}{0.0328} & \multirow{2}{*}{0.0398} & \multirow{2}{*}{0.0041} & \multirow{2}{*}{$8.1190$} & TMM & 1275 & 202 & $7.85\cdot10^{-4}$\\
& & & & & & FEM & 1282 & 204 & $3.43\cdot 10^{-3}$\\
\cline{2-10}
\multirow{2}{*}{\textcolor{red}{$\blacksquare$}}
& \multirow{2}{*}{0.0484} & \multirow{2}{*}{0.0243} & \multirow{2}{*}{0.0400} & \multirow{2}{*}{0.0042} & \multirow{2}{*}{$9.2003$} & TMM & 1220 & 194 & $4.41\cdot10^{-5}$\\
& & & & & & FEM & 1231 & 196 & $7.9\cdot 10^{-3}$\\
\cline{2-10}
\multirow{3}{*}{{\huge$\bullet$}}
& \multirow{3}{*}{0.0380} & \multirow{3}{*}{0.021} & \multirow{3}{*}{0.03} & \multirow{3}{*}{0.0041} & \multirow{3}{*}{$5.4231$} & TMM & 1765 & 281 & $1.85 \cdot 10^{-5}$\\
& & & & & & FEM & 1775 & 283 & $5.43\cdot10^{-3}$\\
& & & & & & EXP & 1791 & 285 & $4.73\cdot10^{-2}$\\
\cline{2-10}
\end{tabular}
\caption{Geometric and acoustic parameters of the optimal resonators selected numerical testing (red symbol rows) and for 3D printing (black circle symbol row). Comparison of the absorption peaks and the resonance frequencies by each method and experimentally for the 3D printed sample.}
\label{tab:values}
\end{table*}

\subsubsection{Neural network}
Regarding the neural network (NN), training is performed using a learning rate $\ell_r=2\cdot 10^{-3}$ and the Adam optimizer, that is an optimization method based on the gradient descent (see \cite{Adam}). Furthermore, to carry out this update, the network observes a batch of 64 experiences at each step. The network takes as input a state $\s \in \R^{n_p}$ and consists of two hidden layers with ReLU activation functions, both with $n_{\text{neu}}$ neurons. The output, without any activation function, is a vector of length $2\cdot n_p + 1$ containing the values of $Q\left(\s,\,a\right)$, where $a$ is one of the actions described earlier. We point out that the choice of using two hidden layers and having the same number of neurons in each layer is arbitrary. 
Specifically, experiments were conducted with
\begin{equation}
    n_{\text{neu}} \in \left\{128,\, 264,\, 512\right\},
\end{equation}
and, although in all cases the resulting fronts reconstructed the same surface, it was decided to perform the final experiments with $n_{\text{neu}} = 512$. The choice of the number of neurons per layer was motivated primarily by the fact that, on average, it yielded a larger number of non-dominated solutions. Furthermore, larger network architectures were not explored due to the computational cost of training and because a lower number of parameters can improve the statistical reliability of the model. The neural network was implemented using the modules provided by PyTorch.

\subsubsection{Rewards}
\label{sub_rewards}
The proposed reward strategy aims to favour configurations with a volume of $V \leq 10^{-4}\,\text{m}^3$, an angular resonance frequency of $\omega_r \leq 2000 \, \text{rad/s}$, and a reflection coefficient modulus of $\abs{R} \leq 0.2$, as these ranges are considered desirable for the optimisation process. Furthermore, greater rewards are assigned to reductions in the absorption coefficient, with the aim of maximizing the number of resonator designs exhibiting near-perfect absorption. This region of interest is motivated by potential application requirements. 

The total reward depends on the decrease and/or increase of the quantities $V$, $\omega_r$, and $\abs{R}$. A reward is assigned to the change in each quantity according to the values reported in Table~\ref{tab:rewards}.

At each step, the total reward is computed from the values of $\mathcal{R}_V$, $\mathcal{R}_{\omega_r}$, and $\mathcal{R}_R$ as follows:
\begin{equation}
\mathcal{R} =
\begin{cases}
-1, & \text{if } \mathcal{R}_V = \mathcal{R}_{\omega_r} = \mathcal{R}_R = 0,\\[4pt]
\mathcal{R}_V + \mathcal{R}_{\omega_r} + \mathcal{R}_R, & \text{otherwise}.
\end{cases}
\label{eq:reward}
\end{equation}
For example, if, at a given step, the volume decreases, the resonance frequency increases, and the reflection coefficient remains unchanged, the resulting reward is $5 - 4 + 0 = 1$. Moreover, if none of the quantities changes, a little penalty is assigned to the selected action in order to discourage stagnation. 
As near-perfect absorbers are desired, a reduction in the reflection coefficient is prioritized over volume or resonance frequency reduction.

In practice, a value is considered constant not when it remains exactly unchanged, but when the following condition is satisfied:
\begin{equation}
    \abs{\dfrac{J^{\text{old}} - J^{\text{new}}}{J^{\text{old}}}} < 10^{-4},
    \label{no_change}
\end{equation}
where $J = V,\, \omega_r,\, \abs{R}$. This condition may occur when the null action is selected, when the enforcement of constraints neutralizes the effect of the selected action, or when the relative change in the magnitude of the variable under consideration is negligible. To prevent the algorithm from becoming stuck repeatedly executing actions that produce no effective change, early episode termination is triggered when all three objective functions remain unchanged, according to Eq.~\eqref{no_change}, for 10 consecutive iterations.

We note that, typically, the reward used in multi-objective problems is assigned through a vector-valued function (see \cite{Hayes2022} and \cite{VanMoffaert2014}). However, to simplify the matter, we decided to consider a scalar reward.

\section{Results}

\subsection{Results of the RL algorthm interpreted as a multi-objective optimization problem}
\label{tuning}
The extraction of the Pareto front was carried out using the \texttt{paretoset} routine from PyPI in Python. This routine selects the non-dominated points through a series of pairwise comparisons. The scan is performed efficiently, minimizing the number of comparisons as much as possible. For further details on the algorithm, see \cite{paretoset}.

Due to the choice of the reward function, the solutions of the Pareto front are concentrated in the region where $V\leq 10^{-4}\,\mathrm{m}$, $\omega_r \leq 2000\,\mathrm{rad/s}$, and $\abs{R} \leq 0.02$. Therefore, it can be assumed that, within this region, the numerically obtained Pareto front provides a good approximation of the theoretical Pareto front and that the corresponding resonators are, therefore, effectively optimal. 

\begin{figure*}
     \centering
     \includegraphics[width=17cm]{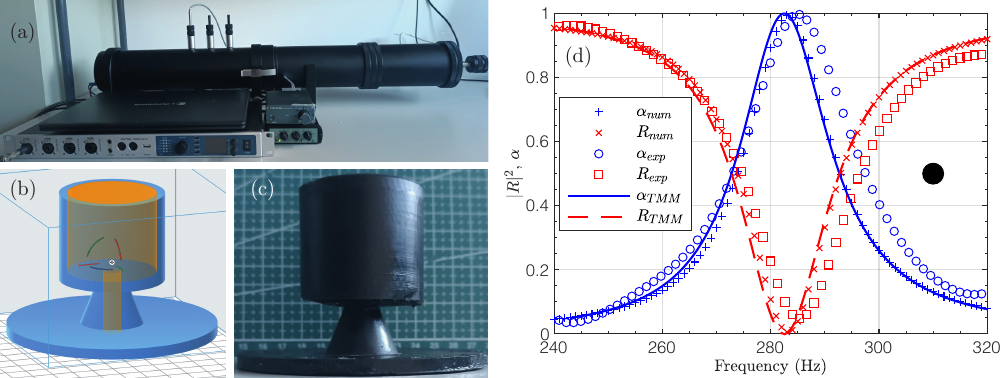}
     \caption{(a) Picture and details of the experimental setup. (b) and (c) show the numerical design of the resonator and the 3D printed sample respectively. (d) shows the absorption (blue) and square of the amplitude f the reflection coefficient (red) for TMM (lines) and full wave simulations bades on FEM (cross symbols) compared with the experimental results (open symbols). The black dor in (d) is a referene to link the results to Tab.~\ref{tab:values}.}
     \label{fig:fig4}
 \end{figure*} 

In Figure \ref{fig:fronte_3D}(a), the Pareto front is shown. We notice that all the points in the Pareto front present very small values of the reflection coefficient, therefore we chose the logarithmic values of the reflection for representation, so we enhance the visibility of the Pareto front.  

To assess the performance of the RL-based optimization framework, four representative designs were selected from the Pareto front and their acoustic absorption coefficients were evaluated. These resonators are highlighted in the Pareto front with big symbols. Three designs were considered to compare the predictions obtained using the TMM with full-wave numerical simulations based on the finite element methods, thereby assessing the accuracy and reliability of the reduced-order model used during the optimization process. The fourth design was selected for experimental validation: the resonator was fabricated using 3D printing and characterized in an impedance tube. This multilevel validation strategy enables the numerical predictions to be assessed against both an independent numerical approach and experimental measurements, providing a comprehensive evaluation of the proposed design methodology. The geometric parameters of the selected resonators can be seen in Tab.~\ref{tab:values}.

\subsection{Validation of the TMM Predictions}
\label{experimental_results}
To validate the effectiveness of the TMM approach with the viscothermal losses, the reflection and absorption spectra of three resonators belonging to the Pareto front [see red symbols (square, pentagram and lozenge) in Fig.~\ref{fig:fronte_3D}(a)] obtained through reinforcement learning were evaluated by the TMM and compared with those obtained using the full wave simulations based on FEM. The results can be seen in Fig.~\ref{fig:fronte_3D}(b-d). Blue (red) line and blue open circles (red open square) show the absorption, $\alpha$, (square of the amplitude of the reflection coefficient, $|R|^2$,) evaluated by the TMM and the full wave simulations. Both approaches exhibit very similar spectral trends for the three samples, with good agreement in the magnitude and frequency dependence of the absorption and reflection coefficients. This consistency validates the TMM as a reliable and computationally efficient approach for modelling the acoustic response of the samples as well as the optimization procedure based on the proposed RL-algorithm. The small frequency shifts observed between the two methods may be attributed to uncertainties in estimating the end correction length in the TMM model, which affects the effective acoustic length of the resonant structures and, consequently, their resonance frequencies.

\subsection{Experimental results}
One of the three resonators was also fabricated using 3D printing, and its reflection coefficient, $R\left(f\right)$, and absorption coefficient, $\alpha\left(f\right)$, were experimentally measured as functions of the excitation frequency, $f = \dfrac{\omega}{2\pi}$. The resonance frequency can also be experimentally determined, as it corresponds to the peak of the absorption coefficient (or, equivalently, to the minimum in the magnitude of the reflection coefficient). 
 
The experimental characterization was performed using a Brüel \& Kjær impedance tube with an inner diameter of 10 cm. The resonator was mounted at the end of the tube and excited using a 20-s swept-sine signal spanning the frequency range from 240 to 320 Hz. The acoustic pressure was measured using a 1/4-inch Brüel \& Kjær Type 2670 microphone, connected to a Brüel \& Kjær Type 1708 signal conditioner. The excitation and acquisition of the acoustic signals were performed using an RME Fireface UFX II audio interface. The normal-incidence sound absorption coefficient was determined using a single-microphone transfer-function method, following the approach described by Chu \cite{SongBolton2000}. In this method, the acoustic pressure is measured sequentially at different positions along the tube to characterize the incident and reflected wave fields and determine the reflection coefficient, from which the absorption coefficient is calculated. Details of the experiemntal setup can be seen in Fig.~\ref{fig:fig4}(a).

The resonator samples were fabricated using Black Resin V4.1 and a Formlabs Form 3 3D printer. To enhance the mechanical rigidity of the structure, the neck was designed with an external conical geometry, while the inner channel through which the acoustic waves propagate retains a cylindrical cross-section. This design strategy increases the structural robustness of the neck while preserving the intended acoustic geometry. Further details of the fabricated resonator are shown in Fig.~\ref{fig:fig4}(b,c).

The geometric and acoustic parameters of the selected resonators are reported in Table \ref{tab:values}, which compares the values obtained using the TMM (including losses and end corrections) and the FEM, together with the experimental measurements for one of the resonators. Furthermore, Fig.~\ref{fig:fig4}(d) compares the three methods in terms of the calculated and measured absorption coefficient, $\alpha\left(f\right)$, and the squared magnitude of the reflection coefficient, $\abs{R\left(f\right)}^2$, as a function of frequency, $f$.

\section{Conclusions}
In this work, a reinforcement learning framework has been developed for the multi-objective design of Helmholtz resonators targeting near-perfect acoustic absorption while controlling their volume and resonance frequency. By formulating the design process as a sequential decision-making problem, the proposed Q-learning algorithm explores the geometrical parameter space and identifies a set of non-dominated configurations that represent the trade-offs between the competing design objectives. The resulting Pareto front provides a systematic means of selecting resonator geometries according to specific acoustic performance and compactness requirements.

The reliability of the optimization framework has been assessed through a combination of reduced-order modelling, full-wave numerical simulations, and experimental measurements. The TMM, including viscothermal losses and acoustic end corrections, reproduces the absorption and reflection spectra obtained from finite element simulations for three representative designs, with only minor shifts in the predicted resonance frequencies. Furthermore, experimental characterization of a three-dimensionally printed resonator in an impedance tube confirms its ability to achieve approximately 95\% acoustic absorption near resonance. The remaining discrepancies between predictions and measurements may be associated with uncertainties in the effective acoustic length, fabrication tolerances, and deviations between the idealized model and the experimental realization.

Overall, these results establish reinforcement learning as a promising tool for the automated design of compact resonant acoustic absorbers. The proposed framework combines the flexibility of data-driven optimization with physically informed acoustic modelling, without requiring gradient information or a pre-existing labelled dataset. Future developments could extend the approach to broadband absorption, multiple resonance frequencies, and more complex geometries, while incorporating manufacturing constraints and experimental feedback directly into the optimization loop.

\section*{Data availability}
The Python codes used in this work are publicly available in the GitHub repository given in  \cite{codes}. Other data shown in the article will be made available on request.

\section*{Acknowledgments}
MN and VRG are funded by the European Union’s Horizon programme in the framework of the DN-MSCA ``ABHSSYS'' project under grant agreement No. 101227712. LMGR, and VRG acknowledge grant PID2023-146237NB-I00 funded by MICIU/AEI/ 10.13039/501100011033. 

\bibliographystyle{unsrt} 



\begin{thebibliography}{10}

\bibitem{Yariv2000Universal}
Amnon Yariv.
\newblock Universal relations for coupling of optical power between
  microresonators and dielectric waveguides.
\newblock {\em Electronics Letters}, 36(4):321--322, 2000.

\bibitem{Chong2010Coherent}
Y.~D. Chong, Li~Ge, Hui Cao, and A.~D. Stone.
\newblock Coherent perfect absorbers: Time-reversed lasers.
\newblock {\em Physical Review Letters}, 105(5):053901, 2010.

\bibitem{RomeroGarcia2016Perfect}
V.~Romero-Garc{\'i}a, G.~Theocharis, O.~Richoux, A.~Merkel, V.~Tournat, and
  V.~Pagneux.
\newblock Perfect and broadband acoustic absorption by critically coupled
  sub-wavelength resonators.
\newblock {\em Scientific Reports}, 6:19519, 2016.

\bibitem{RomeroGarcia2016ComplexFrequency}
V.~Romero-Garc{\'i}a, G.~Theocharis, O.~Richoux, and V.~Pagneux.
\newblock Use of complex frequency plane to design broadband and sub-wavelength
  absorbers.
\newblock {\em The Journal of the Acoustical Society of America},
  139(6):3395--3403, 2016.

\bibitem{RomeroGarcia2020Design}
V.~Romero-Garc{\'i}a, N.~Jim{\'e}nez, G.~Theocharis, V.~Achilleos, A.~Merkel,
  O.~Richoux, V.~Tournat, J.-P. Groby, and V.~Pagneux.
\newblock Design of acoustic metamaterials made of {Helmholtz} resonators for
  perfect absorption by using the complex frequency plane.
\newblock {\em Comptes Rendus Physique}, 21(7--8):713--749, 2020.

\bibitem{Long2017Perfect}
H.~Long, Y.~Cheng, J.~Tao, and X.~Liu.
\newblock Perfect absorption of low-frequency sound waves by critically coupled
  subwavelength resonant system.
\newblock {\em Applied Physics Letters}, 110(2):023502, 2017.

\bibitem{Achilleos2017NonHermitian}
V.~Achilleos, G.~Theocharis, O.~Richoux, and V.~Pagneux.
\newblock Non-hermitian acoustic metamaterials: Role of exceptional points in
  sound absorption.
\newblock {\em Physical Review B}, 95(14):144303, 2017.

\bibitem{Qu2022AbsorptionMetamaterials}
Sichao Qu and Ping Sheng.
\newblock Microwave and acoustic absorption metamaterials.
\newblock {\em Physical Review Applied}, 17:047001, 2022.

\bibitem{Merkel2015Control}
A.~Merkel, G.~Theocharis, O.~Richoux, V.~Romero-Garc{\'i}a, and V.~Pagneux.
\newblock Control of acoustic absorption in one-dimensional scattering by
  resonant scatterers.
\newblock {\em Applied Physics Letters}, 107(24):244102, 2015.

\bibitem{Jimenez2016Ultrathin}
No{\'e} Jim{\'e}nez, Weichun Huang, Vicent Romero-Garc{\'i}a, Vincent Pagneux,
  and Jean-Philippe Groby.
\newblock Ultra-thin metamaterial for perfect and quasi-omnidirectional sound
  absorption.
\newblock {\em Applied Physics Letters}, 109(12):121902, 2016.

\bibitem{LiAssouar2016}
Yong Li and Badreddine~M. Assouar.
\newblock Acoustic metasurface-based perfect absorber with deep subwavelength
  thickness.
\newblock {\em Applied Physics Letters}, 108(6):063502, 2016.

\bibitem{ZhangHu2016SinglePort}
C.~Zhang and X.~Hu.
\newblock Three-dimensional single-port labyrinthine acoustic metamaterial:
  Perfect absorption with large bandwidth and tunability.
\newblock {\em Physical Review Applied}, 6(6):064025, 2016.

\bibitem{DampedResonance2019}
Damped resonance for broadband acoustic absorption in one-port and two-port
  systems.
\newblock {\em Scientific Reports}, 2019.

\bibitem{Jimenez2018PerfectAbsorption}
No{\'e} Jim{\'e}nez, Vicent Romero-Garc{\'i}a, and Jean-Philippe Groby.
\newblock Perfect absorption of sound by rigidly-backed high-porous materials.
\newblock {\em Acta Acustica united with Acustica}, 104(3):396--409, 2018.

\bibitem{Cavalieri2020Graded}
Th{\'e}o Cavalieri, Jean Boulvert, Gw{\'e}na{\"e}l Gabard, Vicent
  Romero-Garc{\'i}a, Marie Escouflaire, Josselin Regnard, and Jean-Philippe
  Groby.
\newblock Graded and anisotropic porous materials for broadband and angular
  maximal acoustic absorption.
\newblock {\em Materials}, 13(20):4605, 2020.

\bibitem{Boulvert2020Folded}
Jean Boulvert, Josu{\'e} Costa-Baptista, Th{\'e}o Cavalieri, Vicent
  Romero-Garc{\'i}a, Gw{\'e}na{\"e}l Gabard, Edith~Roland Fotsing, Annie Ross,
  Maxime Perna, Jacky Mardjono, and Jean-Philippe Groby.
\newblock Folded metaporous material for sub-wavelength and broadband perfect
  sound absorption.
\newblock {\em Applied Physics Letters}, 117(25):251902, 2020.

\bibitem{JimenezSciReps2017}
No{\'e} Jim{\'e}nez, Vicent Romero-Garc{\'i}a, Vincent Pagneux, and
  Jean-Philippe Groby.
\newblock Rainbow-trapping absorbers: Broadband, perfect and asymmetric sound
  absorption by subwavelength panels for transmission problems.
\newblock {\em Scientific Reports}, 7:13595, 2017.

\bibitem{Boulvert2022Asymmetric}
Jean Boulvert, Thomas Humbert, Vicente Romero-Garc{\'i}a, Gw{\'e}na{\"e}l
  Gabard, Edith~Roland Fotsing, Annie Ross, Jacky Mardjono, and Jean-Philippe
  Groby.
\newblock Perfect, broadband, and sub-wavelength absorption with asymmetric
  absorbers: Realization for duct acoustics with 3d printed porous resonators.
\newblock {\em Journal of Sound and Vibration}, 523:116687, 2022.

\bibitem{Boulvert2022Compact}
Jean Boulvert, Gw{\'e}na{\"e}l Gabard, Vicent Romero-Garc{\'i}a, and
  Jean-Philippe Groby.
\newblock Compact resonant systems for perfect and broadband sound absorption
  in wide waveguides in transmission problems.
\newblock {\em Scientific Reports}, 12:10013, 2022.

\bibitem{Meng2023Subwavelength}
Yang Meng, Vicente Romero-Garc{\'i}a, Gw{\'e}na{\"e}l Gabard, Jean-Philippe
  Groby, Charlie Bricault, and S{\'e}bastien Goud{\'e}.
\newblock Subwavelength broadband perfect absorption for unidimensional
  open-duct problems.
\newblock {\em Advanced Materials Technologies}, 8(12):2201909, 2023.

\bibitem{Wei2014CoherentAbsorption}
Pengjiang Wei, Charles Cro{\"e}nne, Sai~Tak Chu, and Jensen Li.
\newblock Symmetrical and anti-symmetrical coherent perfect absorption for
  acoustic waves.
\newblock {\em Applied Physics Letters}, 104(12):121902, 2014.

\bibitem{RomeroGarcia2020MirrorSymmetric}
V.~Romero-Garc{\'i}a, N.~Jim{\'e}nez, J.-P. Groby, A.~Merkel, V.~Tournat,
  G.~Theocharis, O.~Richoux, and V.~Pagneux.
\newblock Perfect absorption in mirror-symmetric acoustic metascreens.
\newblock {\em Physical Review Applied}, 14(5):054055, 2020.

\bibitem{Yang2015DegenerateResonators}
Min Yang, Chong Meng, Caixing Fu, Yong Li, Zhiyu Yang, and Ping Sheng.
\newblock Subwavelength total acoustic absorption with degenerate resonators.
\newblock {\em Applied Physics Letters}, 107(10):104104, 2015.

\bibitem{Ingard1953AcousticResonators}
Uno Ingard.
\newblock On the theory and design of acoustic resonators.
\newblock {\em The Journal of the Acoustical Society of America},
  25(6):1037--1061, 1953.

\bibitem{Li2026PhysicallyAdmissible}
Xinwei Li, Chenxu Liu, Zhendong Li, Miao Zhao, Yiman Chen, Zhonggang Wang, and
  Wei Zhai.
\newblock Navigating competing constraints in acoustic metamaterials via
  physically admissible design spaces.
\newblock {\em Small Structures}, 2026.

\bibitem{Andersen2019ShapeOptimization}
Peter~Risby Andersen, Vicente Cutanda~Henr{\'i}quez, and Niels Aage.
\newblock Shape optimization of micro-acoustic devices including viscous and
  thermal losses.
\newblock {\em Journal of Sound and Vibration}, 2019.

\bibitem{Tissot2020OptimalCavity}
Gilles Tissot, Robin Billard, and Gw{\'e}na{\"e}l Gabard.
\newblock Optimal cavity shape design for acoustic liners using helmholtz
  equation with visco-thermal losses.
\newblock {\em Journal of Computational Physics}, 2020.

\bibitem{Shah2021Reinforcement}
Tristan Shah, Linwei Zhuo, Peter Lai, Amaris De~La Rosa-Moreno, Feruza
  Amirkulova, and Peter Gerstoft.
\newblock Reinforcement learning applied to metamaterial design.
\newblock {\em The Journal of the Acoustical Society of America},
  150(1):321--338, 2021.

\bibitem{Gebrekidan2024Autonomous}
Semere~B. Gebrekidan and Steffen Marburg.
\newblock Autonomous design of noise-mitigating structures using deep
  reinforcement learning.
\newblock {\em The Journal of the Acoustical Society of America},
  156(1):151--163, 2024.

\bibitem{Oh2026UnderwaterCoatings}
Kee~Seung Oh, Cheeyoung Joh, and Joo~Hwan Oh.
\newblock Design of underwater coatings for broadband acoustic absorption using
  reinforcement learning-based topology optimization.
\newblock {\em The Journal of the Acoustical Society of America},
  159(2):1057--1070, 2026.

\bibitem{Lu2026HelmholtzPorous}
Jia-Hao Lu, Siqi Ding, Kai Zhou, and Yi-Qing Ni.
\newblock Dual-learning framework: Integrating deep and reinforcement learning
  for optimizing bio-inspired helmholtz-porous acoustic metamaterials for
  railway noise mitigation.
\newblock {\em Mechanical Systems and Signal Processing}, page 114137, 2026.

\bibitem{Wang2027ReinforcementLearning}
Yinchu Wang, Xiaofeng Li, Mi~Xiao, and Liang Gao.
\newblock Reinforcement learning-driven inverse design of programmable resonant
  acoustic metamaterials.
\newblock {\em Thin-Walled Structures}, 232(Part 1):115757, 2027.

\bibitem{Jimenez2021TMM}
No{\'e} Jim{\'e}nez, Jean-Philippe Groby, and Vicent Romero-Garc{\'i}a.
\newblock The transfer matrix method in acoustics.
\newblock In No{\'e} Jim{\'e}nez, Olga Umnova, and Jean-Philippe Groby,
  editors, {\em Acoustic Waves in Periodic Structures, Metamaterials, and
  Porous Media: From Fundamentals to Industrial Applications}, pages 103--164.
  Springer, 2021.

\bibitem{Stinson1991}
Michael~R. Stinson.
\newblock The propagation of plane sound waves in narrow and wide circular
  tubes, and generalization to uniform tubes of arbitrary cross-sectional
  shape.
\newblock {\em The Journal of the Acoustical Society of America},
  89(2):550--558, 1991.

\bibitem{vicent}
N.~Jiménez, O.~Umnova, and J.~P. Groby.
\newblock {\em Acoustic Waves in Periodic Structures, Metamaterials, and Porous
  Media: From Fundamentals to Industrial Applications}.
\newblock Springer, 2021.

\bibitem{Dubos}
V.~Dubos, J.~Kergomard, A.~Khettabi, J.-P. Dalmont, D.~H. Keefe, and C.~J.
  Nederveen.
\newblock Theory of sound propagation in a duct with a branched tube using
  modal decomposition.
\newblock {\em Acta Acustica united with Acustica}, 85(2):153--169, 1999.

\bibitem{Dalmont1998Measurement}
Jean-Pierre Dalmont and Cornelis~J. Nederveen.
\newblock Measurement of the radiation impedance of a pipe with a circular
  flange.
\newblock {\em The Journal of the Acoustical Society of America}, 103(5), 1998.

\bibitem{Dalmont2001Radiation}
Jean-Pierre Dalmont, Cornelis~J. Nederveen, and Nicolas Joly.
\newblock Radiation impedance of tubes with different flanges: numerical and
  experimental investigations.
\newblock {\em Journal of Sound and Vibration}, 244(3):505--534, 2001.

\bibitem{Gebr2024}
S.~B. Gebrekidan and S.~Marburg.
\newblock Autonomous design of noise-mitigating structures using deep
  reinforcement learning.
\newblock {\em The Journal of the Acoustical Society of America}, 156:151--163,
  July 2024.

\bibitem{Mitchell1997}
T.~M. Mitchell.
\newblock {\em Machine Learning}.
\newblock McGraw-Hill, New York, 1997.

\bibitem{Yu20}
S.~Yu, X.~Piao, and N.~Park.
\newblock Machine learning identifies scale-free properties in disordered
  materials.
\newblock {\em Nature Communications}, 11(1):4842, 2020.

\bibitem{SuttonBarto2018}
R.~S. Sutton and A.~G. Barto.
\newblock {\em Reinforcement Learning: An Introduction}.
\newblock MIT Press, Cambridge, MA, 2nd edition, 2018.

\bibitem{Zhao2024}
S.~Zhao.
\newblock {\em Mathematical Foundations of Reinforcement Learning}.
\newblock Springer, 2025.

\bibitem{Hayes2022}
C.~F. Hayes, R.~Rădulescu, E.~Bargiacchi, J~Källström, M.~Macfarlane,
  M.~Reymond, T.~Verstraeten, L.~M. Zintgraf, R.~Dazeley, F.~Heintz, E.~Howley,
  A.~A. Irissappane, P.~Mannion, A.~Nowé, G.~Ramos, M.~Restelli, P.~Vamplew,
  and D.~M. Roijers.
\newblock A practical guide to multi-objective reinforcement learning and
  planning.
\newblock {\em Journal of Autonomous Agents and Multi-Agent Systems}, 36(26),
  2022.
\newblock Published online: 13 April 2022.

\bibitem{VanMoffaert2014}
K.~Van Moffaert and A.~Now{\'e}.
\newblock Multi-objective reinforcement learning using sets of pareto
  dominating policies.
\newblock {\em Journal of Machine Learning Research}, 15:3663--3692, Nov 2014.
\newblock Submitted 12/13; Revised 7/14; Published 11/14.

\bibitem{Watkins1989}
C.~J. C.~H. Watkins.
\newblock {\em Learning from Delayed Rewards}.
\newblock Ph.d. thesis, King's College, University of Cambridge, Cambridge, UK,
  1989.

\bibitem{Mnih2013}
V.~Mnih, K.~Kavukcuoglu, D.~Silver, A.~Graves, I.~Antonoglou, D.~Wierstra, and
  M.~Riedmiller.
\newblock Playing atari with deep reinforcement learning.
\newblock {\em arXiv preprint arXiv:1312.5602}, 2013.

\bibitem{Arnau2023}
R.~Arnau, L.~M. García-Raffi, J.~M. Calabuig, and E.~A.~Sánchez Pérez.
\newblock Aprendizaje por refuerzo para minimizar funciones.
\newblock In {\em IX Congreso de Innovación Educativa y Docencia en Red},
  Valencia, España, julio 2023.

\bibitem{Bellman1957}
R.~E. Bellman.
\newblock {\em Dynamic Programming}.
\newblock Princeton University Press, Princeton, NJ, 1957.

\bibitem{Adam}
D.~P. Kingma and J.~L. Ba.
\newblock Adam: A method for stochastic optimization.
\newblock In {\em Proceedings of the 3rd International Conference on Learning
  Representations (ICLR)}, 2015.
\newblock Published as a conference paper at ICLR 2015.

\bibitem{paretoset}
S.~Börzsönyi, D.~Kossmann, and K.~Stocker.
\newblock The skyline operator.
\newblock In {\em Proceedings of the 17th International Conference on Data
  Engineering (ICDE)}, pages 421--430. IEEE, 2001.

\bibitem{SongBolton2000}
Bryan~H. Song and J.~Stuart Bolton.
\newblock A transfer-matrix approach for estimating the characteristic
  impedance and wave numbers of limp and rigid porous materials.
\newblock {\em The Journal of the Acoustical Society of America},
  107(3):1131--1152, 2000.

\bibitem{codes}
M.~Novelli, R.A. Arnau~Notari, L.M. García-Raffi, and V.~Romero-García.
\newblock {Reinforcement learning codes}.
\newblock
  \url{https://github.com/Mattianovelli/Reinforcement-learning-to-design-near-perfect-acoustic-absorbers},
  2026.
\newblock GitHub repository, Version 1.0.

\end{thebibliography}





\end{document}